\documentclass{new_tlp}
\usepackage{mathptmx}
\usepackage{amsmath}
\usepackage{amssymb}

\newtheorem{theorem}{Theorem}
\newtheorem{lemma}{Lemma}
\newcommand*{\qed}{\hfill\ensuremath{\square}}%
\def\proof{\noindent{\bf Proof}\hspace{3mm}}

\def\pnn{\hbox{\rm Pnn}}
\def\nnn{\hbox{\rm Nnn}}
\def\Int{\hbox{\rm int}}
\def\I{\mathcal{I}}
\def\o{\overline}
\def\ar{\leftarrow}
\def\rar{\rightarrow}
\def\lrar{\leftrightarrow}
\def\no{\i{not}}
\def\i#1{\hbox{\it #1\/}}

\def\beq{\begin{equation}}
\def\eeq#1{\label{#1}\end{equation}}
\def\ba{\begin{array}}
\def\ea{\end{array}}

\title[Program Completion in GRINGO]{Program Completion\\
in the Input Language of GRINGO}

\author[Harrison, Lifschitz, and Raju]{AMELIA HARRISON,   
VLADIMIR LIFSCHITZ \and DHANANJAY RAJU
\\
University of Texas
\\ \email{ameliaj,vl,draju@cs.utexas.edu}
}

\begin{document}

\maketitle 

\begin{abstract}
We argue that turning a logic program into a set of completed definitions
can be sometimes thought of as the ``reverse engineering'' process of
generating a set of conditions that could serve as a specification for it.
Accordingly, it may be useful to define completion for a large class of ASP
programs and to automate the process of generating and simplifying
completion formulas.  Examining the output produced
by this kind of software may help programmers to see more clearly what their
program does, and to what degree its behavior conforms with their
expectations.  As a step toward this goal, we propose here
a definition of program completion for a large class of programs in the
input language of the ASP grounder {\sc gringo}, and study its properties. 
This paper is published in Theory and Practice of Logic Programming, Volume 17, 
Issue 05-06 (the special issue on the 32nd International Conference on Logic Programming).

\end{abstract}

\section{Introduction}

Our interest in defining completion \cite{cla78} for programs in the input
language of the ASP grounder {\sc gringo} ({\tt https://potassco.org}) is
motivated by the goal of extending formal methods
for software verification to answer set programming.  Turning a logic
program into a set of completed definitions can be sometimes thought of as the
``reverse engineering'' process of generating a set of conditions that
could serve as a specification for it.  Consider, for instance, the condition
``set~$r$ is the union of sets~$p$ and~$q$.''
In the language of logic programming this definition of~$r$ is represented
by the pair of rules
\beq
\ba l
r(X) \ar p(X),\\
r(X) \ar q(X).
\ea
\eeq{u1}
The corresponding completed definition
$$
\forall X(r(X) \lrar p(X)\lor q(X))
$$
is the usual definition of union in set theory. Turning program~(\ref{u1})
into a completed definition gives us a plausible specification that could
have led to this program in the first place.  The stable model semantics
of program~(\ref{u1}) matches the completed definition, because the
program is tight \cite{fag94,erd03}.

It may be useful to define completion for a large class of ASP programs and
to automate the process of generating and simplifying completion formulas.
(Simplifying is essential because ``raw'' completion rarely provides such a
clean specification as in the example above.)  Examining the output produced
by this kind of software may help programmers to see more clearly what their
program does, and to what degree its behavior conforms with their
expectations.  If the programming project started with a formal
specification then they may be able to verify the correctness of the
program relative to that specification by comparing the given specification
with the ``engineered specification'' extracted from the program.

As a step toward this goal, we propose here a definition of program
completion for a large class of {\sc gringo} programs.
Three issues need to be addressed.  First, {\sc gringo} programs often
include constraints and choice rules, which are not covered by
Clark's theory.  Extending completion to these constructs has been
discussed in the literature; see, for instance, \cite[Section~6.1]{fer09}.

Second, we need to take into account the fact that in the language of
{\sc gringo} a ground term may denote a set of values, rather
than a single value.  For instance, the term $1..8$ denotes the set
$\{1,\dots,8\}$, and the condition $X=1..8$ in the body of a rule expresses
that $X$ is an element of that set.  In standard mathematical notation, this
condition would be expressed using the set membership symbol rather than
equality.  The syntax of {\sc gringo} allows us to write also
$$X..X\!+\!1=Y..Y\!+\!1,$$
which is understood as
$$\hbox{$X$ and $Y$ are integers, and }
\{X,X\!+\!1\}\cap\{Y,Y\!+\!1\}\neq\emptyset.$$

Third, the semantics of aggregate expressions in the language of {\sc gringo}
depends on the distinction between local and global variables.  This is
similar to the distinction between bound and free variables familiar from
first-order logic, except that the definition of a local variable does not
refer to quantifiers.  The expression
$\i{sum}\{X\!\times\! Y\,:\, p(X,Y)\}$
in the body of a rule\footnote{We use here an ``abstract'' syntax, which
disregards some details related to writing rules as strings of ASCII characters
\cite[Section~1]{geb15}.  In an actual {\sc gringo} program this expression
would be written as {\tt \#sum\{X*Y:p(X,Y)\}}.}
may correspond to any of the expressions
$$\sum_{X,Y\,:\,p(X,Y)}X\!\times\! Y,\quad\sum_{X\,:\,p(X,Y)}X\!\times\! Y,
\quad\sum_{Y\,:\,p(X,Y)}X\!\times\! Y
$$
depending on where $X$ and $Y$ occur in other parts of the rule.
Our way of translating aggregate expressions takes into account
this feature.  Otherwise it is similar to the approach proposed
by Ferraris and Lifschitz~\citeyear{fer10a}, which is closely related to
the use of generalized quantifiers by Lee and 
Meng~(\citeyearNP{lee09}, \citeyearNP{lee12a}).  One of their results
\cite[Theorem~4]{lee12a} relates stable models of formulas with
generalized quantifiers to program completion.

We start by discussing a class of programs that do not contain aggregate
expressions.   Sections~\ref{sec:programs} and~\ref{sec:formulas} define a
language of programs and a language of formulas---the source and the
target of the completion operator.  Section~\ref{sec:representing}
describes the process of representing rules by formulas, which is
used in the definition of completion in Section~\ref{sec:es}.  We discuss
tight programs in Section~\ref{sec:tight} and give an example of
calculating an engineered specification in Section~\ref{sec:example}.
Incorporating aggregate expressions is described in Section~\ref{sec:incorp}.
In Section~\ref{sec:intvar} the class of formulas is further extended by
adding variables for integers, which can be often used to simplify
formulas that involve arithmetic operations.
The definition of a stable model for the class of programs defined in
Section~\ref{sec:incorp} is given in Appendix~A.  Proofs of theorems are
given in Appendix~B. 

\section{Programs}\label{sec:programs}

We assume that four disjoint sets of symbols are selected:
{\sl numerals}; {\sl symbolic constants}; {\sl variables}; and
{\sl operation names} of various arities.
We assume that these sets do not contain the {\sl interval symbol}
$$..$$
the {\sl relation symbols}
$$= \qquad \not = \qquad <\qquad >\qquad \leq\qquad \geq$$
and the symbols
$$
\i{inf}\qquad \i{sup}\qquad \no\qquad\land\qquad \lor\qquad\ar
$$
$$
,\qquad ;\qquad :\qquad (\qquad )\qquad \{\qquad \}
$$
$$
\in\qquad \neg\qquad \land\qquad \lor\qquad\rar\qquad\lrar\qquad
\forall\qquad\exists$$

We assume that a 1--1 correspondence between the set of numerals and the
set~${\bf Z}$ of integers is chosen.  For every integer~$n$, the
corresponding numeral will be denoted by~$\o n$.
We will identify a numeral with the corresponding integer when this does
not lead to confusion.

We assume that for every operation name $\i{op}$, a function
$\widehat{\i{op}}$
from a subset of ${\bf Z}^n$ to~${\bf Z}$ is chosen, where~$n$ is the arity
of \i{op}.  For instance, we can choose \i{plus} as a binary operation
name, define $\widehat{\i{plus}}$ as the addition of integers, and use
$t_1\!+\!t_2$ as shorthand for $\i{plus}(t_1,t_2)$.

{\sl Terms} are defined recursively, as follows:
\begin{itemize}
\item numerals, symbolic constants, variables, and the symbols \i{inf}
and \i{sup} are terms,
\item if~$f$ is a symbolic constant and~${\bf t}$ is a non-empty tuple
of terms (separated by commas) then $f({\bf t})$ is a term,
\item if \i{op} is an $n$-ary operation name and~${\bf t}$ is an $n$-tuple
of terms then $\i{op}({\bf t})$ is a term,
\item if $t_1$ and $t_2$ are terms then $(t_1..t_2)$ is a term.
\end{itemize}

A term, or another syntactic expression, is {\sl ground} if it does not
contain variables.  A ground term is
{\sl precomputed} if it contains neither operation names nor the
interval symbol.
According to the semantics of terms defined in Section~A.1,
every ground term~$t$ denotes a finite set $[t]$ of precomputed terms, which
are called the {\sl values} of~$t$.  For instance,
$$[\o 8]=\{\o 8\},\ [\o 1..\o 8]=\{\o 1,\dots,\o 8\},\ [\i{abc}+\o 1]=\emptyset$$
if \i{abc} is a symbolic constant.  

We assume a total order on precomputed terms such that \i{inf} is its least
element, \i{sup} is its greatest element, and,
for any integers $m$ and $n$, $\o m \leq \o n$ iff $m \leq n$.  

{\sl Atoms} are expressions of the form $p({\bf t})$, where $p$ is a
symbolic constant and ${\bf t}$ is a tuple of terms, possibly empty.
An atom of the form $p()$ will be written as~$p$.
{\sl Literals} are atoms ({\sl positive literals\/}) and atoms preceded
by~$\no$ ({\sl negative literals\/}).
A {\sl comparison} is an expression of the form $(t_1\prec t_2)$
where~$t_1$,~$t_2$ are terms and $\prec$ is a relation symbol.

A {\sl choice expression} is an expression of the form $\{A\}$ where~$A$ is an
atom.

A {\sl rule} is an expression of the form
\beq
\i{Head}\ar\i{Body}
\eeq{rule}
where
\begin{itemize}
\item
{\it Body} is a conjunction (possibly empty) of literals and comparisons, and
\item
{\it Head} is either an atom (then
we say that~(\ref{rule}) is a {\sl basic rule\/}), or a choice expression
(then~(\ref{rule}) is a {\sl choice rule}\/), or empty (then~(\ref{rule})
is a {\sl constraint}).
\end{itemize}
If the body of a basic rule or choice rule is empty then the arrow will
be dropped.

A {\sl program} is a set of rules.

An {\sl interpretation} is a set of atoms of the form
$p({\bf t})$ where $\bf t$ is a tuple of precomputed terms.
Every program denotes a set of interpretations, which are called its
{\sl stable models} (Appendix~A).

\section{Formulas}\label{sec:formulas}

The language defined in this section is essentially
a first-order language with variables for precomputed terms.

An {\sl argument} is a term that contains neither operation names nor the
interval symbol.\footnote{Thus precomputed terms (Section~\ref{sec:programs})
can be alternatively described as ground arguments.  This will not be the
case, however, when we extend the definition of an argument in
Section~\ref{ssec:fwag} to incorporate aggregates.}
{\sl Formulas} are defined recursively:
\begin{itemize}
\item[(a)] if~$p$ is a symbolic constant and {\bf arg} is a tuple
of arguments then $p({\bf arg})$ is a formula,
\item[(b)] if $\i{arg}_1$ and $\i{arg}_2$ are arguments and $\prec$ is a
relation symbol then $(\i{arg}_1\prec\i{arg}_2)$ is a formula,
\item[(c)] if \i{arg} is an argument and~$t$ is a term then $\i{arg}\in t$
is a formula,
\item[(d)] $\bot$ (``false'') is a formula,
\item[(e)] if $F$ and $G$ are formulas then $(F\rar G)$ is a formula;
\item[(f)] if $F$ is a formula and $X$ is a variable then $\forall XF$
is a formula.
\end{itemize}
We will drop parentheses in formulas when it does not lead to confusion.
Propositional connectives other than implication, and the existential
quantifier, are defined as abbreviations in the usual way.
Free and bound occurrences of variables, closed formulas, and the
universal closure of a formula are defined as usual in first-order logic.

Note that a term that is not an argument can occur in a formula in only one
position---to the right of the $\in$ symbol.  For example, $X\in \o 1..\o 8$
and $X\in Y\!+\!\o 1$ are formulas, but $X=\o 1..\o 8$ and $X=Y\!+\!\o 1$
are not.
The reason why we do not allow $Y\!+\!\o 1$ in equalities is that
substituting a precomputed term for~$Y$ in this expression (for
instance, \i{abc}) may give a term that has no values.

If $F$ is a formula, $X$ is a variable, and~$r$ is a precomputed term, then
$F^X_r$ stands for the formula obtained from~$F$ by substituting $r$ for
all free occurrences of~$X$.

The truth value $F^\I$, assigned by an interpretation~$\I$ to a closed
formula~$F$, is defined as~{\sf t} or~{\sf f}, in accordance with the
following rules:
\begin{itemize}
\item[(a)] $p({\bf arg})^\I$ is {\sf t} if $p({\bf arg})\in \I$ (and~{\sf f}
otherwise),
\item[(b)] $(\i{arg}_1\prec \i{arg}_2)^\I$ is {\sf t}
if $\i{arg}_1 \prec \i{arg}_2$,
\item[(c)] $(\i{arg}\in t)^\I$ is {\sf t} if $\i{arg}\in[t]$,
\item[(d)] $\bot^\I$ is {\sf f},
\item[(e)] $(F\rar G)^\I$ is {\sf f} if $F^\I$ is {\sf t} and $G^\I$ is {\sf f},
\item[(f)] $(\forall XF)^\I$ is {\sf t} if, for every precomputed term~$r$,
$(F^X_r)^\I$ is {\sf t}.
\end{itemize}

We say that an interpretation~$\I$ {\sl satisfies} a closed formula~$F$ if
$F^\I={\sf t}$.

For example, the interpretation $\{p(\o 2),p(\o 3),p(\o 4)\}$ satisfies
the formula
$
\exists X (p(X)\land X\in \o 1..\o 8).
$
Indeed, it satisfies $p(\o 3)$, because it includes~$p(\o 3)$; it also
satisfies $\o 3\in \o 1..\o 8$, because $[\o 1..\o 8]$ is
$\{\o 1,\dots, \o 8\}$, and~$\o 3$
is an element of this set.  Consequently it satisfies
the conjunction $p(\o 3)\land \o 3\in \o 1..\o 8$.

A formula is {\sl universally valid} if its universal closure is satisfied
by all interpretations.
A formula~$F$ is {\sl equivalent} to a formula~$G$ if $F\lrar G$ is
universally valid.  Since our definition of satisfaction treats
propositional connectives,
quantifiers, and equality in the same way as the standard definition of
satisfaction applied to the domain of precomputed terms, all equivalent
transformations sanctioned by classical first-order logic can be used
in this setting as well.  The following additional observations about
equivalence will be useful.

\medskip\noindent{\bf Observation~1.}  {\sl For any argument $\i{arg}$ and
any ground term~$t$, $\i{arg}\in t$ is equivalent to
\hbox{$\bigvee_{r\in [t]}(\i{arg}=r)$.}}

\medskip
This is immediate from the definition of satisfaction.

For example, for any integers $m$ and $n$, $\i{arg}\in \o m..\o n$ is
equivalent to
$\bigvee_{i=m}^n(\i{arg}=\o i)$.

\medskip\noindent{\bf Observation~2.}  {\sl For any arguments $\i{arg}_1$
and $\i{arg}_2$, $\i{arg}_1\in\i{arg}_2$ is equivalent to
$\i{arg}_1=\i{arg}_2$.}

\medskip
It is sufficient to check this claim for the case when $\i{arg}_1$,
$\i{arg}_2$ are ground.  In this case, it follows from the fact
that $[\i{arg}_2]$ is the singleton $\{\i{arg}_2\}$.

For example, $X\in Y$ is equivalent to $X=Y$.

\section{Representing Rules by Formulas}\label{sec:representing}

In this section we define a syntactic transformation $\phi$ that turns rules
and their subexpressions into formulas---their {\sl formula representations}.

Formula representations of literals and comparisons are defined as follows:
\begin{itemize}
\item
$\phi\, p({\bf t})$ is
$\exists {\bf X} 
({\bf X} \in {\bf t} \land p({\bf X}))$,\footnote{If $\bf X$ is
$X_1,\dots,X_n$, and $\bf t$ is $t_1,\dots,t_n$, then
${\bf X} \in {\bf t}$ stands for the conjunction
$\bigwedge_{i=1}^n X_i \in t_i.$}
\item
$\phi(\no\ p({\bf t}))$ is
$\exists {\bf X} 
({\bf X} \in {\bf t} \land \neg p({\bf X}))$,
\item
$\phi(t_1\prec t_2)$ is
$\exists X_1X_2(X_1\in t_1 \land X_2\in t_2 \land X_1\prec X_2)$;
\end{itemize}
here ${\bf X}$ is a tuple of new variables of the same length as ${\bf t}$, and
$X_1, X_2$ are new variables.

For example, the transformation~$\phi$ turns $p(X)$ into
$\exists Y(Y\in X \land p(Y))$;
this formula is equivalent to $\exists Y(Y=X \land p(Y))$, and consequently
to $p(X)$.  The formula representation of $p(\o 1..X)$ is
$\exists Y(Y\in \o 1..X \land p(Y))$.  The representation of $X=\o 1..\o 8$ is
$$\exists X_1X_2(X_1\in X\land X_2\in \o 1..\o 8\land X_1=X_2);$$
this formula is equivalent to $X\in \o 1..\o 8$.

If each of the expressions $C_1,\dots, C_k$ is a literal or a comparison then
$\phi(C_1\land \cdots \land C_k)$ stands for 
$\phi C_1\land \cdots \land  \phi C_k.$

The formula representation of a basic rule
\beq
p({\bf t}) \ar\i{Body}
\eeq{basicrule}
is defined as the implication
\beq
{\bf V} \in {\bf t} \land \phi(\i{Body}) \rar
  p({\bf V}),
\eeq{bfr}
where~${\bf V}$ is a tuple of new variables of the same length as ${\bf t}$.
For example, the formula representation of the rule
\beq
q(X\!+\!\o 1) \ar p(X) \land X=\o 1..\o 8
\eeq{qrule}
is
$$V\in X\!+\!\o 1 \land \phi\,p(X) \land \phi\,(X=\o 1..\o 8)\,\rar\, q(V);$$
after applying equivalent transformations to the
antecedent, this formula becomes
\beq
V\in X\!+\!\o 1 \land p(X) \land X\in \o 1..\o 8 \,\rar\, q(V).
\eeq{simpant}

The formula representation of a choice rule
\beq
\{p({\bf t})\} \ar \i{Body}
\eeq{choicerule}
is defined as the (universally valid) formula
\beq
{\bf V} \in {\bf t} \land \phi(\i{Body})
\land p({\bf V}) \rar p({\bf V}),
\eeq{cfr}
where ${\bf V}$ is a tuple of new variables of the same length as ${\bf t}$.

For example, the formula representation of the rule
$\{p(\o 1..\o 8)\}$
is
$$
V\in \o 1..\o 8 \land p(V) \rar p(V).
$$

The formula representation of a constraint $\ar\i{Body}$ is the formula
\beq
\neg\phi(\i{Body}).
\eeq{consfr}

\section{Completion}\label{sec:es}

A {\sl predicate symbol} is a pair $p/n$, where~$p$ is a symbolic
constant and $n$ is a nonnegative integer.
The {\sl definition} of a predicate symbol $p/n$ in a 
program~$\Gamma$ consists of
\begin{itemize}
\item the basic rules of~$\Gamma$ with the head of the form
$p(t_1,\dots,t_n)$, and
\item the choice rules of~$\Gamma$ with the head of the form
$\{p(t_1,\dots,t_n)\}$.
\end{itemize}
It is clear that any program is the union of the
definitions of predicate symbols and a set of constraints.

If the definition of~$p/n$ in a finite program~$\Gamma$ is
$\{R_1,\dots,R_k\}$ then each of the formulas $\phi R_i$ has the form
\beq
F_i\rar p({\bf V}),
\eeq{imp}
where~$\bf V$ is a tuple of distinct variables.
We will assume that this tuple is chosen in the same way for all~$i$.
The {\sl completed definition} of $p/n$ in~$\Gamma$ is the formula
\beq
\forall {\bf V}\left(p({\bf V}) \lrar 
                             \bigvee_{i=1}^k\exists {\bf U}_i F_i\right),
\eeq{compdef}
where ${\bf U}_i$ is the list of all free variables of the formula
$F_i$ that do not belong to~{\bf V}.

For example if the definition of~$p/1$ in~$\Gamma$ is $p(\o 1..\o 8)$ then
$k=1$, ${\bf U}_1$ is empty, and $F_1$ is $V\in \o 1..\o 8$, so that
the completed definition of~$p/1$ is
\beq
\forall V (p(V) \lrar V \in \o 1..\o 8).
\eeq{cdex1}
If the definition of~$p/1$ is the choice rule $\{p(\o 1..\o 8)\}$ then~$F_1$
is
$$V\in \o 1..\o 8 \land p(V),$$
and the completed definition of~$p/1$ is
$$
\forall V(p(V) \lrar V\in \o 1..\o 8\land p(V)).
$$
This formula is equivalent to
\beq
\forall V(p(V) \rar V\in \o 1..\o 8).
\eeq{cdex2}

It is clear that completed definitions are invariant with respect to
equivalent transformations of the antecedents of implications $\phi R_i$, in
the sense that replacing an antecedent~$F_i$ in~(\ref{compdef}) by an
equivalent formula is an equivalent transformation.
Assume, for instance, that the definition of~$q/1$ in~$\Gamma$
is~(\ref{qrule}).  Formula~(\ref{simpant}) is the result of
simplifying the antecedent of the formula representation of that rule, and
the completed definition of~$q/1$ can be written as
\beq
\forall V(q(V) \lrar\exists X(V\in X\!+\!\o 1 \land p(X)
                                         \land X\in \o 1..\o 8)).
\eeq{cdex3}

About a program or another
syntactic expression we say that a predicate symbol $p/n$ {\sl
occurs} in it if it contains an atom of the form $p(t_1,\dots,t_n)$.
The {\sl completion} of a finite program~$\Gamma$ consists of
\begin{itemize}
\item the completed definitions of all predicate symbols occurring
in~$\Gamma$, and
\item the universal closures of the formula representations of all
constraints in~$\Gamma$.
\end{itemize}

The definition of completion matches the stable model semantics in the
following sense:

\begin{theorem}
Every stable model of a finite program satisfies its completion.
\label{thm1}
\end{theorem}

In the next section we define a class of programs for which the converse of
Theorem~1 can be proved.

\section{Tight Programs}\label{sec:tight}

For any program~$\Gamma$, by $G_\Gamma$ we denote the directed graph that
has the predicate symbols occurring in~$\Gamma$ as its vertices, and
has an edge from $q/m$ to $p/n$ if $\Gamma$ includes a rule~$R$ such that
\begin{itemize}
\item[(i)] $q/m$ occurs in the head of~$R$, and
\item[(ii)] $p/n$ occurs in a positive literal in the body of~$R$.
\end{itemize}
If graph~$G_\Gamma$ is acyclic then we will say that program~$\Gamma$ is
{\sl tight}.

Consider, for instance, the program $\Gamma_{r,n}$ ($r$ and $n$ are
positive integers) that consists of the rules
\begin{align}
 &\{\i{in}(1..\o n,1..\o r)\},\label{r1}\\
 \i{covered}(X) &\ar \i{in}(X,S),\label{r2}\\
 &\ar X=1..\o n \land \no\ \i{covered}(X),\label{r3}\\
 &\ar \i{in}(X,S)\land \i{in}(Y,S)\land \i{in}(X\!+\!Y,S).\label{r4}
\end{align}
(The stable models of 
this program represent collections of~$r$ sum-free sets covering
$\{1,\dots,n\}$; see {\tt http://mathworld.wolfram.com/SchurNumber.html}.)
The graph $G_{\Gamma_{r,n}}$ has one edge, from $\i{covered}/1$ to
$\i{in}/2$, so that this program is tight.

The {\sl vocabulary} of a program~$\Gamma$, is the set of atoms
$p({\bf r})$ such that $\bf r$ is a tuple of~$n$ precomputed terms,
and $p/n$ occurs in~$\Gamma$.  For other syntactic expressions the
vocabulary is defined in the same way.

\begin{theorem}
For any tight finite program~$\Gamma$, an interpretation~$\I$ is a stable model
of~$\Gamma$ iff~$\I$ is contained in the vocabulary of~$\Gamma$ and satisfies
the completion of~$\Gamma$.
\label{thm2}
\end{theorem}

The theorem shows, for instance, that the stable models
of $\Gamma_{r,n}$ can be characterized as the subsets of its vocabulary that
satisfy its completion.

\section{Example}\label{sec:example}

We will now calculate and simplify the completion of~$\Gamma_{r,n}$.
The formula representation of rule~(\ref{r1}) is
$$V_1\in \o 1..\o n \land V_2\in \o 1..\o r \land\i{in}(V_1,V_2)\,\rar\,
\i{in}(V_1,V_2),$$
so that the completed definition of $\i{in}/2$ is
$$\forall V_1V_2(\i{in}(V_1,V_2) \lrar
(V_1\in \o 1..\o n \land V_2\in \o 1..\o r \land \i{in}(V_1,V_2))).$$
This formula is equivalent to
\beq
\forall V_1V_2(\i{in}(V_1,V_2) \rar
(V_1\in \o 1..\o n \land V_2\in \o 1..\o r)).
\eeq{compin}

The formula representation of rule~(\ref{r2}) can be written as
$$V=X \land \i{in}(X,S) \rar \i{covered}(V).$$
It follows that the completed definition of $\i{covered}/1$ is
$$\forall V(\i{covered}(V) \lrar \exists XS(V=X \land \i{in}(X,S))),$$
which is equivalent to
\beq
\forall V(\i{covered}(V) \lrar \exists S\,\i{in}(V,S)).
\eeq{compcovered}

The remaining two rules of the program are constraints.  The universal
closure of the formula representation of~(\ref{r3}) is equivalent to
$$\forall X\neg(X\in\o 1..\o n \land\neg\i{covered}(X)),$$
which can be further rewritten as
\beq
\forall X(X\in\o 1..\o n \rar\i{covered}(X)).
\eeq{c1}

Finally, the universal closure of the formula representation of
constraint~(\ref{r4}) can be written as
\beq
\neg\exists XYS(\i{in}(X,S)\land\i{in}(Y,S)\land
\exists Z(Z\in X\!+\!Y\land \i{in}(Z,S))).
\eeq{c2}
We showed that the completion of program~$\Gamma_{r,n}$---its ``engineered
specification''---is equivalent to the
conjunction of formulas (\ref{compin})--(\ref{c2}).

\section{Incorporating Aggregates}\label{sec:incorp}

\subsection{Programs with Aggregates} \label{ssec:pwag}

In addition to the four sets of symbols mentioned at the beginning of
Section~\ref{sec:programs}, we assume now that a set of {\sl aggregate names}
is selected, and for every aggregate name~$\alpha$ a function
$\widehat \alpha$ is chosen that maps every set of non-empty tuples of
precomputed terms to a precomputed term.  Examples:
\begin{itemize}
\item
aggregate name {\it count}; $\widehat{\i{count}}(T)$ is 
defined as the cardinality of~$T$ if~$T$ is
finite, and $\i{sup}$ otherwise;
\item
aggregate name {\it sum}; $\widehat{\i{sum}}(T)$ is the sum of the
weights of all tuples in~$T$ if~$T$ contains finitely many tuples with
non-zero weights, and $\o 0$ otherwise.
\end{itemize}
(The {\sl weight} of a tuple~${\bf t}$ of precomputed terms is
the first member of~${\bf t}$ if it is a numeral, and $\o 0$ otherwise.)

An {\sl aggregate expression} is an expression of the form
\beq
\alpha\{{\bf t} : {\bf C} \} \prec s
\eeq{agex0}
where $\alpha$ is an aggregate name, ${\bf t}$ is a non-empty tuple of terms,
${\bf C}$ is a conjunction of literals and comparisons (in the case when 
${\bf C}$ is  empty the preceding colon can be dropped),
$\prec$ is a relation symbol, and $s$ is a variable or precomputed term.

In the definition of a rule, the body is now allowed to have, among its
conjunctive terms, not only literals and comparisons, but also aggregate
expressions.

A variable~$V$ occurring in a rule~$R$ is {\sl local} if every occurrence
of~$V$ in~$R$ belongs to the left-hand side $\alpha\{{\bf t} : {\bf C} \}$
of one of the aggregate expressions~(\ref{agex0}) in its body, and
{\sl global} otherwise.  For instance, in the rule
\beq
q(W) \ar \i{sum}\{X^2 : p(X)\}=W
\eeq{rulewa}
$X$ is local and~$W$ is global.

\subsection{Formulas with Aggregates} \label{ssec:fwag}

The definitions of an argument and a formula in Section~\ref{sec:formulas}
are replaced now by a mutually recursive definition of both concepts.  It
includes clauses (a)--(f) from the old definition of a formula and three
additional clauses:
\begin{itemize}
\item[(g)]
numerals, symbolic constants, variables, and the symbols \i{inf}
and \i{sup} are arguments; 
\item[(h)]
if~$f$ is a symbolic constant and~${\bf arg}$ is a non-empty tuple
of arguments then $f({\bf arg})$ is an argument;
\item[(i)]
if $\alpha$ is an aggregate name, {\bf X} is a non-empty tuple of
distinct variables, and $F$ is a formula, then $\alpha\{{\bf X}\,|\,F\}$
is an argument.
\end{itemize}
Clause (i) is what makes the new definition more general than the
definitions from Section~\ref{sec:formulas}.

In this more general setting, the distinction between free and bound
occurrences of variables applies not only to formulas, but also to arguments.
An occurrence of a variable~$X$ in an argument or in a formula is
{\sl bound} if it belongs to a subformula of the form $\forall XF$, or if
it belongs to a subargument $\alpha\{{\bf X}\,|\,F\}$
such that~$X$ is a member of the tuple {\bf X}.  For example, in the argument
$$\i{sum}\{X\,|\,\exists Y p(X,Y,Z)\}$$
$X$ and $Y$ are bound, and~$Z$ is free.  An argument or a
formula is {\sl closed} if all occurrences of variables in it are bound.

The substitution notation will be now applied not only to formulas, but also
to arguments: $\i{arg}^X_r$ is the argument obtained from an
argument~$\i{arg}$ by substituting a precomputed term~$r$ for all free
occurrences of a variable~$X$.
For every interpretation~$\I$, the truth value $F^\I$ that~$\I$ assigns to a
closed formula~$F$, and the precomputed term $\i{arg}^\I$ that~$\I$ assigns
to a closed argument \i{arg}, are described by a joint recursive
definition:\footnote{This notation can be ambiguous,
because some expressions can be viewed both as formulas and as arguments.
But its meaning will be always clear from the context.}
\begin{itemize}
\item[(a)] $p(\i{arg}_1,\dots,\i{arg}_k)^\I$ is {\sf t} if
$p(\i{arg}_1^\I,\dots,\i{arg}_k^\I)\in \I$,
\item[(b)] $(\i{arg}_1\prec\i{arg}_2)^\I$ is {\sf t} if
$\i{arg}_1^\I \prec \i{arg}_2^\I$,
\item[(c)] $(\i{arg}\in t)^\I$ is {\sf t} if $\i{arg}^\I\in[t]$,
\item[(d)] $\bot^\I$ is {\sf f},
\item[(e)] $(F\rar G)^\I$ is {\sf f} if $F^\I$ is {\sf t} and $G^\I$ is {\sf f},
\item[(f)] $(\forall XF)^\I$ is {\sf t} if, for every precomputed term~$r$,
$(F^X_r)^\I$ is {\sf t}.
\item[(g)] if~\i{arg} is a numeral, or a symbolic constant, or~\i{inf},
or~\i{sup}, then $\i{arg}^\I$ is \i{arg};
\item[(h)] $f(\i{arg}_1,\dots,\i{arg}_k)^\I$ is
$f(\i{arg}_1^\I,\dots,\i{arg}_k^\I)$,
\item[(i)] $\alpha\{X_1,\cdots,X_k\,|\,F\}^\I$ is $\widehat\alpha(T)$,
where~$T$ is the set of all tuples $r_1,\dots,r_k$ of precomputed terms
such that $(F^{X_1\cdots X_k}_{\;\,r_1\,\dots\,r_k})^\I$ is {\sf t}.
\end{itemize}

Since an argument containing aggregate names is not a term, in this more
general setting the statement of Observation~2 (Section~\ref{sec:formulas})
has to be modified:

\medskip\noindent{\bf Observation~2$'$.}  {\sl For any arguments $\i{arg}_1$
and $\i{arg}_2$ such that $\i{arg}_2$ does not contain aggregate names,
$\i{arg}_1\in\i{arg}_2$ is equivalent to $\i{arg}_1=\i{arg}_2$.}

\subsection{Completion and Tightness in the Presence of Aggregates}
\label{ssec:tightaggr}

How do we turn an aggregate expression \eqref{agex0}
into a formula?  It depends on how we classify the variables occurring in
this expression into local and global.  For this reason, instead of
extending the definition of~$\phi$ from Section~\ref{sec:representing}
to aggregate expressions, we will
define the transformation~$\phi^{\bf X}$, where $\bf X$ is a list (possibly
empty) of distinct variables---those that we treat as local.  The
result of applying~$\phi^{\bf X}$ to an aggregate expression~(\ref{agex0})
is the formula
$$\exists Y(\alpha\{{\bf Z}\,|\,
\exists {\bf X}({\bf Z}\in {\bf t} \land\phi {\bf C})
\}\prec Y\land Y\in s),$$
where ${\bf Z}$ is a tuple of new variables of the same length as ${\bf t}$, and 
$Y$ is a new variable.

Consider, for instance, the result of applying the transformation
$\phi^X$ (``treat~$X$ as local'') to the aggregate expression in the body of
rule~(\ref{rulewa}).  It can be written as
$$\exists Y(\i{sum}\{ Z\,|\,\exists X(Z\in X^2 \land p(X))\}=Y\land Y=W),$$
which is equivalent to
$$\i{sum}\{ Z\,|\,\exists X(Z\in X^2 \land p(X))\}=W.$$

In application to literals and comparisons, $\phi^{\bf X}$ has the same
meaning as~$\phi$.  If each of the expressions $C_1,\dots, C_k$ is a
literal, a comparison, or an aggregate expression, then
\hbox{$\phi^{\bf X}(C_1\land \cdots \land C_k)$} stands for
$\phi^{\bf X} C_1\land \cdots \land \phi^{\bf X} C_k.$

Now we are ready to state how the definitions~(\ref{bfr}),~(\ref{cfr}),
and~(\ref{consfr}) of formula representations of rules are
modified in the presence of aggregates.  In all three definitions, we replace
$\phi(\i{Body})$ by $\phi^{\bf X}(\i{Body})$, where~$\bf X$ is the list of local
variables of the rule.  For instance, the formula representation of
rule~(\ref{rulewa}) can be written as
$$V=W \land \i{sum}\{ Z\,|\,\exists X(Z\in X^2 \land p(X))\}=W \rar q(V).$$

All definitions from Section~\ref{sec:es}, including the definition of
the completion of a finite program, remain the same.  It is easy to see
that in formula~(\ref{compdef}), ${\bf U}_i$
is the list of global variables of rule~$R_i$.

In the definition of~$G_\Gamma$ (Section~\ref{sec:tight}), clause~(ii) is
restated as follows:
\begin{itemize}
\item[(ii$'$)] $p/n$ occurs in a positive literal or in an aggregate
expression in the body of~$R$.
\end{itemize}
For example, if~$\Gamma$ is the one-rule program~(\ref{rulewa}) then
$G_\Gamma$ has an edge from $q/1$ to $p/1$.  Otherwise, the definition of a
tight program remains the same.

\subsection{Example: $8$-Queens}

The following program with aggregates encodes a solution to the problem of
how to place 8 queens on an $8\times 8$ chessboard so that no two queens
attack each other.   
\begin{align}
&\i{row}(\o 1..\o 8),\label{q1}\\
&\i{col}(\o 1..\o 8),\label{q2}\\                                   
&\{ \i{queen}(X,Y) \} \ar \i{col}(X) \land \i{row}(Y),\label{q3a}\\
&\ar \i{count}\{X,Y : \i{queen}(X,Y) \}\neq\o 8,\label{q3b}\\
&\ar \i{queen}(X,Y)\land \i{queen}(X,\i{YY})\land Y \neq \i{YY},
                                                            \label{q4}\\
&\ar \i{queen}(X,Y)\land \i{queen}(\i{XX},Y)\land X \neq \i{XX},
                                                            \label{q5}\\
&\ar \i{queen}(X,Y)\land \i{queen}(\i{XX},\i{YY})\land
                  X \neq \i{XX} \land |X-\i{XX}|=|Y-\i{YY}|.\label{q6}
\end{align}
The formula representation of rule~(\ref{q1}) is
$V\in \o 1..\o 8 \rar \i{row}(V)$,
so that the completed definition of $\i{row}/1$ is
\beq
\forall V(\i{row}(V) \lrar V\in \o 1..\o 8).
\eeq{cd1}
Similarly, the completed definition of $\i{col}/1$ is
\beq
\forall V(\i{col}(V) \lrar V\in \o 1..\o 8).
\eeq{cd2}

The formula representation of~(\ref{q3a}) can be rewritten, after simplifying
the antecedent, as
$$
V_1=X \land V_2=Y \land \i{col}(X) \land \i{row}(Y)
                \land \i{queen}(V_1,V_2) \rar \i{queen}(V_1,V_2).
$$
Consequently the completed definition of \i{queen}/2 is
$$
\forall V_1V_2(\i{queen}(V_1,V_2) \lrar
  \exists XY(V_1=X \land V_2=Y \land \i{col}(X) \land \i{row}(Y)
                              \land \i{queen}(V_1,V_2))).
$$
This formula is equivalent to
\beq
\forall V_1V_2(\i{queen}(V_1,V_2) \rar \i{col}(V_1) \land \i{row}(V_2)).
\eeq{cd3}

Variables $X$ and $Y$ are local in constraint~\eqref{q3b}, so that its
formula representation can be written as
$$
\exists Y_1(\i{count}\{Z_1,Z_2\,|\,\exists XY(Z_1\in X\land Z_2\in Y
\land \i{queen}(X,Y) )\}\neq Y_1 \land Y _1=\o 8)\rar\bot,
$$
or, equivalently,
\beq
\i{count}\{Z_1,Z_2\,|\,\i{queen}(Z_1,Z_2) \} =\o 8.
\eeq{qc}

The formula representations of constraints~(\ref{q4})--(\ref{q6}) can be
written as 
\beq
\ba l
\i{queen}(X,Y)\land \i{queen}(X,\i{YY})\rar Y = \i{YY},\\
\i{queen}(X,Y)\land \i{queen}(\i{XX},Y)\rar X = \i{XX},\\
\i{queen}(X,Y)\land \i{queen}(\i{XX},\i{YY})\land |X-\i{XX}|=|Y-\i{YY}|
                  \rar X = \i{XX}.
\ea\eeq{f456}

The completion of program~(\ref{q1})--(\ref{q6}) consists of
formulas~(\ref{cd1})--(\ref{qc})  and
the universal closures of formulas~(\ref{f456}).
This set of formulas is an ``engineered specification'' for that
program.

\section{Integer Variables} \label{sec:intvar}

We will now make the definition of formulas and arguments more general.
We assume here that the set of variables is partitioned into two classes,
{\sl general variables} and {\sl integer variables}.  General variables are
variables for precomputed terms; integer variables are variables for
numerals.  Formulas without general variables are similar to formulas of
first-order arithmetic.  In examples,
integer variables will be represented by identifiers that start with
$I$, $J$, $K$, $L$, $M$, and $N$.

{\sl Integer arguments} are defined recursively:
\begin{itemize}
\item numerals and integer variables are integer arguments;
\item if \i{op} is an $n$-ary operation name such that the domain of
the corresponding function $\widehat{\i{op}}$ is the whole set ${\bf Z}^n$,
and~${\bf arg}$ is an $n$-tuple of integer arguments, then
$\i{op}({\bf arg})$ is an integer argument.
\end{itemize}
Clause~(g) in the definition of formulas and arguments
(Section~\ref{ssec:fwag}) is reformulated as follows:
\begin{itemize}
\item[(g)]
integer arguments, symbolic constants, general variables,
\i{inf}, and \i{sup} are arguments.
\end{itemize}

For example, since~$N$ is an integer variable, the expression $N\!+\!\o 1$ is
not only a term but also an argument, and both $p(N\!+\!\o 1)$ and
$N\!+\!\o 1=\o 4$ are formulas.

To extend the definition of the semantics of formulas and arguments given in
Section~\ref{ssec:fwag}, we restrict clause~(f) in that definition to the
case when~$X$ is a general variable, and add two clauses:
\begin{itemize}
\item[(f$'$)] $(\forall NF)^\I$, where~$N$ is an integer variable, is {\sf t}
if, for every integer~$n$, $(F^X_{\o n})^\I$ is {\sf t};
\item[(h$'$)] if \i{arg} is $\i{op}(\i{arg}_1,\dots,\i{arg}_k)$,
$\i{arg}^\I_1=\o{n_1},\dots,\i{arg}^\I_k=\o{n_k}$, then
$\i{arg}^\I$ is $\o{\widehat{\i{op}}(n_1,\dots,n_k)}$.
\end{itemize}

The following abbreviations will be useful.
For any argument~\i{arg}, by $\Int(\i{arg})$ we denote the formula
$\exists V(V\in\i{arg}\!+\!\o 1)$, where~$V$ is a general variable that does not
occur in~\i{arg}.  For any predicate symbol $p/n$, by $\Int(p/n)$ we denote
the formula
$$\forall X_1\dots X_n(p(X_1,\dots,X_n) \,\rar\,
 \Int(X_1) \land\cdots\land \Int(X_n)),$$
where $X_1,\dots,X_n$ are distinct general variables.  This formula
expresses that the extent of the predicate $p/n$ is a subset of ${\bf Z}^n$.

Using integer variables, we can rewrite formula~(\ref{cdex1}) as
$$\ba l
\Int(p/1),\\
\forall N(p(N) \lrar \o 1\leq N\leq \o 8).
\ea$$
Formula~(\ref{cdex2}) can be transformed in a similar way.

Formula~(\ref{cdex3}) can be rewritten as
$$\ba l
\Int(q/1),\\
\forall N(q(N) \lrar \exists M(N=M\!+\!\o 1 \land p(M) \land
                                            \o 1\leq M\leq \o 8)).
\ea$$
The last formula can be simplified as follows:
$$\forall N(q(N) \lrar p(N\!-\!\o 1) \land \o 2\leq N\leq \o 9).$$

Formula~(\ref{compin}) is equivalent to
$$\ba l
\Int(in/2),\\
\forall IK(\i{in}(I,K)\,\rar\,\o 1\leq I\leq \o n \land \o 1\leq K\leq \o r).
\ea
$$
Formula~(\ref{c1}) is equivalent to
$$\forall I(\o 1\leq I\leq \o n \rar \i{covered}(I)).$$
Formula~(\ref{c2}) can be equivalently rewritten, in the presence of
$\Int(in/2)$, as
$$
\neg\exists IJS(\i{in}(I,S)\land\i{in}(J,S)\land
\exists K(K\!=\!I\!+\!J\land \i{in}(K,S))),
$$
and consequently as
$$
\forall IJS(\i{in}(I,S) \land\i{in}(J,S)\rar\neg\i{in}(I\!+\!J,S)).
$$

In the presence of completed definitions~(\ref{cd1})--(\ref{cd3}), all
variables in~(\ref{qc}) and in the universal closures of~(\ref{f456})
can be equivalently replaced by integer variables.

\section{Conclusion}

This paper extends familiar results on the relationship between stable models
and program completion to a large class of programs in the input language
of {\sc gringo}, and we hope that this technical contribution will help us
apply formal methods to answer set programming.  Much still remains to be done.

First, we would like to extend the main result of this paper,
Theorem~\ref{thm2}
from Section~\ref{sec:tight}, in several directions.  Including edges
from head to aggregate expressions in graph~$G_\Gamma$ (condition~(ii$'$)
in Section~\ref{ssec:tightaggr}) may be unnecessary when the aggregates are
known to be monotone or antimonotone \cite[Section~6.1]{har14a}.
Further, a dependency graph with atoms from the program's
vocabulary as its vertices, rather than predicate symbols, may be useful.
Finally, we would like to adapt the definition of completion to a class
of ``almost tight'' programs that may contain simple recursive
definitions (such as the definition of reachability in a graph).  It may
be possible to achieve this at the price of allowing the least fixed point
operator \cite{gur86} in completed definitions.

Second, the process of generating and simplifying completed definitions
needs to be automated.  In some cases, programmers may be able to
convince themselves that a program is correct---or to decide that it
is not---by examining its simplified completion.  Sometimes automated
reasoning tools may help them establish a correspondence between a
given specification and the completion of the program.
These are themes of an ongoing
project\footnote{\tt https://github.com/potassco/anthem/} at the University
of Potsdam, the home of {\sc gringo}.

\section*{Acknowledgements}

This work was partially supported by the National Science Foundation under
Grant IIS-1422455.  Many thanks to Michael Gelfond, Joohyung Lee,
Patrick L\"uhne, and Torsten Schaub for useful discussions related to the
topic of this paper, and to the anonymous referees for their comments.

\appendix

\section{Semantics of Programs}

Gebser~et~al. \citeyear{geb15} showed that stable models of many
programs in the input language of {\sc gringo} can be described in
terms of stable models of infinitary propositional formulas.
That approach is applied here to programs in the sense of
Section~\ref{ssec:pwag}; we will call them {\sl EG programs} (for ``Essential
{\sc gringo}''). 

The translation~$\tau$, defined below, transforms every EG
program~$\Gamma$ into an infinitary formula over the vocabulary of~$\Gamma$.
Stable models of~$\Gamma$ are defined as stable models
of~$\tau\Gamma$.\footnote{The stable model semantics of infinitary formulas
\cite{tru12}, \cite[Section~4.1]{geb15} is a straightforward generalization
of the definition due to Ferraris \citeyear{fer05}.}

\subsection{Semantics of Ground Terms}

The set~$[t]$ of precomputed terms denoted by a ground term~$t$ is defined
recursively:
\begin{itemize}
\item if $t$ is a numeral, a symbolic constant, or one of the symbols
\i{inf}, \i{sup} then $[t]$ is $\{t\}$;
\item if $t$ is $f(t_1, \dots, t_n)$, where $f$ is a symbolic constant, then
$[t]$ consists of the terms $f(r_1, \dots, r_n)$ for all
$r_1 \in [t_1], \dots,r_n \in [t_n]$; 
\item if $t$ is $\i{op}(t_1, \dots, t_n)$ where $\i{op}$ is an operation
name then $[t]$ consists of the numerals of the form
$\o{\widehat{\i{op}}(k_1,\dots,k_n)}$
for all tuples $k_1,\dots,k_n$ in the domain of $\widehat{\i{op}}$
such that $\o{k_1} \in [t_1],\dots$, $\o{k_n} \in [t_n]$;
\item if $t$ is $(t_1\, ..\, t_2)$ then $[t]$ consists of the
numerals~$\o m$ for all integers $m$ such that,
for some integers $k_1, k_2,$  
$$ \o{k_1} \in [t_1], \qquad \o{k_2} \in [t_2], \qquad k_1 \leq m \leq k_2.$$
\end{itemize}

For any ground terms $t_1 \dots, t_n$, $[t_1, \dots, t_n]$ 
is the set of tuples $r_1, \dots, r_n$ for all
$r_1 \in [t_1], \dots,$ $r_n \in [t_n]$.

\subsection{Transforming Programs into Infinitary Formulas}
\label{ssec:trans}

For any ground atom $p({\bf t})$,
$\tau p({\bf t})$ stands for 
$\bigvee_{{\bf r} \in [{\bf t}]} p({\bf r})$, 
and 
$\tau (\no \; p({\bf t}))$ stands for 
$\bigvee_{{\bf r} \in [{\bf t}]} \neg p({\bf r})$.

For any ground comparison $t_1 \prec t_2$, $\tau(t_1 \prec t_2)$ is~$\top$
if the relation~$\prec$ holds between some terms  $r_1$, $r_2$ such that
$r_1 \in [t_1]$ and $r_2 \in [t_2]$, and~$\bot$ otherwise. 

If each of $C_1,\dots, C_k$ is a ground literal
or a ground comparison then $\tau(C_1\land \cdots \land C_k)$
stands for $\tau C_1\land \cdots \land \tau C_k$.

An aggregate expression~(\ref{agex0}) is {\sl closed} if the term~$s$ is
ground.
Let~${\bf X}$ be the list of variables occurring in a closed aggregate
expression~(\ref{agex0}), and
let~$A$ be the set of tuples~${\bf r}$ of precomputed terms of the same
length as~${\bf X}$.  Let $\Delta$ be a subset of $A$. By $[\Delta]$ we 
denote the union of the sets $[{\bf t}^{\bf X}_{\bf r}]$ for all tuples of 
precomputed terms~${\bf r}$ in $\Delta$. We say that~$\Delta$ {\sl justifies} 
the aggregate expression~(\ref{agex0}) if the relation $\prec$ holds between 
$\widehat\alpha[\Delta]$ and~$s$.  We define the result of applying~$\tau$
to~(\ref{agex0}) as the conjunction of the implications
\beq
\bigwedge_{{\bf r}\in\Delta}\tau({\bf C}^{\bf X}_{\bf r})
\rar\bigvee_{{\bf r}\in A\setminus\Delta}\tau({\bf C}^{\bf X}_{\bf r})
\eeq{eq:agtrans}
over all subsets $\Delta$ of $A$ that do not justify~(\ref{agex0}).

The definition of $\tau$ for conjunctions of ground literals
and ground comparisons extends in
the obvious way to the case when some conjunctive terms are closed
aggregate expressions.

A rule is {\sl closed} if all its variables are local. 
If~$R$ is a closed basic rule~(\ref{basicrule})
then $\tau R$ is the formula
\beq
\tau(\i{Body}) \rar
\bigwedge_{{\bf r}\in [{\bf t}]} p({\bf r}).
\eeq{taubasic}
If~$R$ is a closed choice rule~(\ref{choicerule})
then $\tau R$ is the formula
\beq
\tau(\i{Body}) \rar \bigwedge_{{\bf r} \in [{\bf t}]}
(p({\bf r})\lor\neg p({\bf r})).
\eeq{tauchoice}
If~$R$ is a closed constraint $\ar\i{Body}$ then $\tau R$ is
$\neg\tau(\i{Body})$.

An {\sl instance} of a rule is a closed rule obtained from it by
substituting precomputed terms for its global variables.  For any EG
program~$\Gamma$, $\tau\Gamma$ is the conjunction of the formulas~$\tau R$
for all instances~$R$ of the rules of~$\Gamma$.

\section{Proofs}

\subsection{Relationship between $\phi$ and $\tau$}

To prove Theorems~\ref{thm1} and~\ref{thm2}, we need to investigate
the relationship between the operator~$\phi$ used in the definition of
completion (Section~\ref{sec:es}) and the operator~$\tau$ that the
semantics of programs is based on (Section~\ref{ssec:trans}).

If~$\bf C$ is a conjunction of ground literals and ground comparisons then
the formula~$\tau {\bf C}$ is finite, and we can ask whether it is
equivalent to~$\phi {\bf C}$ in the sense of Section~\ref{sec:formulas}.
The answer to this question is yes:

\begin{lemma}
For any conjunction~${\bf C}$ of ground literals and ground comparisons,
$\tau {\bf C}$ is equivalent to~$\phi {\bf C}$.
\label{lem:phitaueasy}
\end{lemma}

\proof  It is sufficient to prove this assertion assuming
that~${\bf C}$ is a single ground literal or a single ground comparison.

\medskip\noindent
{\sl Case~1:} ${\bf C}$ is a ground atom $p(t_1,\dots,t_n)$.  Then
$\phi {\bf C}$ is
$$\exists x_1\dots x_n(x_1 \in t_1 \land\cdots \land x_n\in t_n \land
p(x_1,\dots, x_n)).$$
In view of Observation~1, this formula is equivalent to
$$\exists x_1\dots x_n\left(\left(\bigvee_{r_1\in [t_1]}x_1=r_1\right)
         \land\cdots\land\left(\bigvee_{r_n\in [t_n]}x_n=r_n\right)
         \land p(x_1,\dots, x_n)\right),$$
and consequently to
$$\bigvee_{r_1\in [t_1],\dots,r_n\in [t_n]}p(r_1,\dots,r_n).$$
The last formula is $\tau {\bf C}$.

\medskip\noindent
{\sl Case~2:} ${\bf C}$ is a negative ground literal $\neg p(t_1,\dots,t_n)$.
The proof is similar.

\medskip\noindent
{\sl Case~3:} ${\bf C}$ is a ground comparison $t_1 \prec t_2$.  Then
Then $\phi{\bf C}$ is
$$\exists x_1x_2(x_1 \in t_1 \land x_2\in t_2 \land x_1\prec x_2).$$
In view of Observation~1, this formula is equivalent to
$$\exists x_1x_2\left(\left(\bigvee_{r_1\in [t_1]}x_1=r_1\right)
         \land\left(\bigvee_{r_2\in [t_2]}x_2=r_2\right)
         \land x_1\prec x_2\right),$$
and consequently to
$$\bigvee_{r_1\in [t_1],r_2\in [t_2]}r_1\prec r_2.$$
If the relation~$\prec$ holds between some terms  $r_1$, $r_2$ such that
$r_1 \in [t_1]$ and $r_2 \in [t_2]$ then one of the disjunctive terms
in the last formula is~$\top$, and the formula is equivalent to~$\top$;
otherwise each disjunctive term is~$\bot$, and the formula is
equivalent to~$\bot$.  In both cases, it is equivalent to $\tau {\bf C}$.
\qed

\begin{lemma}\label{lem:altagtrans}
For any closed aggregate expression~$E$ and any list ${\bf X}$ of
distinct variables containing all variables that occur in $E$, the
infinitary formula~$\tau E$ is satisfied by the same interpretations of
the vocabulary of~$E$ as the EG formula~$\phi^{\bf X}E$.
\end{lemma}

\proof
Let~$E$ be a closed aggregate expression~(\ref{agex0}).  Without loss of
generality we can assume that the list~{\bf X} contains only variables
occurring in~$E$.  As defined in Section~\ref{ssec:trans},~$\tau E$
is the conjunction of formulas~\eqref{eq:agtrans}, where~$A$ stands for the
set of tuples of precomputed terms of the same length as~${\bf X}$,
over the subsets $\Delta$ of~$A$ that do not justify~$E$.

Note first that~$\tau E$ is classically equivalent to the
disjunction of formulas
\beq
\bigwedge_{{\bf r} \in \Delta} \tau({\bf C}^{\bf X}_{\bf r}) \land  
\bigwedge_{{\bf r} \in  A\setminus\Delta}\neg\tau({\bf C}^{\bf X}_{\bf r})
\eeq{eq:altagtrans}
over the subsets $\Delta$ of $A$ that justify $E$.  
Indeed, call this disjunction $D^+$, and let $D^-$ be the disjunction of 
formulas \eqref{eq:altagtrans} over all other subsets $\Delta$ of $A$.
It is clear that $D^-$ is classically equivalent to $\neg D^+$; on
the other hand, $\neg D^-$ is classically equivalent to the
conjunction~$\tau E$.

Consider now an interpretation~$\I$ of the vocabulary of~$E$.  Set~$A$ has
exactly one subset~$\Delta$ for which~$\I$ satisfies~(\ref{eq:altagtrans}):
the set of all tuples ${\bf r}$ for
which~$\I\models\tau({\bf C}^{\bf X}_{\bf r})$.  Consequently~$\I$
satisfies~$\tau E$ iff this subset~$\Delta$ justifies~$E$.  In other
words,~$\I$ satisfies~$\tau E$ iff, for some~$s'\in[s]$,
\beq
\widehat\alpha\left(\bigcup_{{\bf r}\,:\,\I\models\tau({\bf C}^{\bf X}_{\bf r})}
[{\bf t}^{\bf X}_{\bf r}]\right)\prec s'.
\eeq{int1}
By Lemma~\ref{lem:phitaueasy}, the condition
$\I\models\tau({\bf C}^{\bf X}_{\bf r})$ in this expression can be
equivalently replaced by  $\I\models\phi({\bf C}^{\bf X}_{\bf r})$, and
consequently by $\I\models(\phi{\bf C})^{\bf X}_{\bf r}$.  Hence~(\ref{int1})
holds iff
\beq
\widehat\alpha\{{\bf q}\,:\,\hbox{there exists {\bf r} such that }
{\bf q}\in[{\bf t}^{\bf X}_{\bf r}]\hbox{ and }
\I\models(\phi{\bf C})^{\bf X}_{\bf r}\}\prec s'.
\eeq{int2}
On the other hand, $\phi^{\bf X}E$ is
$$\exists Y (\alpha\{{\bf Z}\;|\;\exists {\bf X}
   ({\bf Z} \in {\bf t} \land \phi{\bf C})\}\prec Y\land Y\in s),$$
and~$\I$ satisfies this formula iff, for some $s'\in s$,
$$\I\models \alpha\{{\bf Z}\;|\;\exists {\bf X}
   ({\bf Z} \in {\bf t} \land \phi{\bf C})\}\prec s'.$$
This condition can be rewritten as
$$\widehat\alpha\{{\bf q}\,:\,\I\models \exists {\bf X}
   ({\bf q} \in {\bf t} \land \phi{\bf C})\}\prec s',$$
which is equivalent to~(\ref{int2}).
\qed\medskip

From Lemmas~\ref{lem:phitaueasy} and~\ref{lem:altagtrans} we conclude:
\begin{lemma}
For any conjunction~${\bf C}$ of ground literals, ground comparisons,
and closed aggregate expressions, and for any list ${\bf X}$ of
distinct variables containing all variables that occur in {\bf C}, the
infinitary formula~$\tau{\bf C}$ is satisfied by the same interpretations of
the vocabulary of~{\bf C} as the EG formula~$\phi^{\bf X}{\bf C}$.
\label{lem:phitau}
\end{lemma}

\subsection{Relation to Infinitary Programs} \label{ssec:relinf}

An {\sl infinitary rule} is an implication $F\rar A$ such
that~$F$ is an infinitary formula and~$A$ is an atom.  An {\sl infinitary
program} is a conjunction of (possibly infinitely many) infinitary rules.
We will prove Theorems~\ref{thm1} and~\ref{thm2} using properties of infinitary
programs proved by Lifschitz and Yang \citeyear{lif13a}.  The
result of applying transformation~$\tau$ to an EG program is, generally,
not an infinitary program, and the following definitions will be useful.

For any EG program~$\Gamma$, by~$\tau_1\Gamma$ we denote the
conjunction of
\begin{itemize}
\item the infinitary rules
\beq
\tau(\i{Body})\rar p({\bf r})
\eeq{basicinf}
for all instances~(\ref{basicrule})
of the basic rules of~$\Gamma$ and all {\bf r} in $[{\bf t}]$, and
\item the infinitary rules
\beq
\tau(\i{Body})\land \neg\neg p({\bf r})\rar p({\bf r})
\eeq{choiceinf}
for all instances~(\ref{choicerule})
of the choice rules of~$\Gamma$ and all {\bf r} in $[{\bf t}]$.
\end{itemize}
By~$\tau_2\Gamma$ we denote the conjunction
of the infinitary formulas $\neg\tau{\bf C}$ for all instances
$\ar {\bf C}$ of the constraints of~$\Gamma$.

\begin{lemma}
Stable models of an EG program~$\Gamma$ can be characterized as the
stable models of the infinitary program $\tau_1\Gamma$ that satisfy
$\tau_2\Gamma$.
\label{lem:tau12}
\end{lemma}

\proof
 The infinitary formula obtained by applying~$\tau$ to a closed basic
rule~(\ref{basicrule}) is strongly equivalent to the conjunction of
the infinitary rules~(\ref{basicinf}) for all {\bf r} in $[{\bf t}]$,
because these two formulas are equivalent in the deductive
system~$\i{HT}^\infty$ \cite[Section 6]{har15a}.
Similarly, the infinitary formula obtained by applying~$\tau$ to a closed
choice rule~(\ref{choicerule}) is strongly equivalent to the conjunction of
the infinitary rules~(\ref{choiceinf}) for all~{\bf r} in~$[{\bf t}]$.
It follows that~$\Gamma$ has the same stable models as
$\tau_1\Gamma\cup\tau_2\Gamma$.  We know, on the other hand, that for any
infinitary formula~$F$ and any conjunction~$G$ of infinitary formulas
that begin with negation, stable models of $F\land G$ can be characterized
as the stable models of~$F$ that satisfy~$G$.  (This is a
straightforward extension of Proposition~4 from Ferraris and 
Lifschitz~\citeyear{fer05e} to infinitary formulas.)  It remains to apply this
general fact to $\tau_1\Gamma$ as~$F$ and $\tau_2\Gamma$ as~$G$.
\qed\medskip

For any infinitary program~$\Pi$ and any atom~$A$, by $\Pi|_A$ we denote the
set of formulas~$F$ such that $F\rar A$ is a rule of~$\Pi$.
The {\sl completion} of~$\Pi$ is the conjunction of the formulas
$A\lrar (\Pi|_A)^\lor$ for all atoms~$A$ in the underlying signature.

\begin{lemma}
For any finite EG program~$\Gamma$,
the completion of the infinitary program~$\tau_1\Gamma$ is satisfied
by the same interpretations of the vocabulary of~$\Gamma$ as the set of
completed definitions of the predicate symbols occurring in~$\Gamma$.
\label{lem:comp}
\end{lemma}

\proof
We will show, for every predicate symbol~$p/n$ occurring
in~$\Gamma$, that its completed definition~(\ref{compdef}) is satisfied by
the same interpretations of the vocabulary of~$\Gamma$ as the conjunction
of the formulas
$$p({\bf r})\lrar (\tau_1\Gamma|_{p({\bf r})})^\lor$$
over all tuples~{\bf r} of precomputed terms of length~$n$.  An interpretation
satisfies~(\ref{compdef}) iff it satisfies the formulas
$$p({\bf r}) \lrar
    \bigvee_{i=1}^k\exists {\bf U}_i (F_i)^{\bf V}_{\bf r}$$
for all tuples~{\bf r} of precomputed terms of length~$n$.  Consequently
it is sufficient to check that for every such tuple~{\bf r}, the
infinitary formula
\beq
(\tau_1\Gamma|_{p({\bf r})})^\lor
\eeq{dis1}
and the EG formula
\beq
\bigvee_{i=1}^k\exists {\bf U}_i (F_i)^{\bf V}_{\bf r}
\eeq{dis2}
are satisfied by the same interpretations.

The rules of $\tau_1\Gamma$ with the consequent $p({\bf r})$ are obtained as
described in the definition of~$\tau_1$ above from instances of the rules
$R_1,\dots,R_k$ that define~$p/n$ in ~$\Gamma$.  If~$R_i$ is a basic rule
\beq
p({\bf t}_i) \ar \i{Body}_i
\eeq{bi}
then its instances have the form
$$
p\left(({\bf t}_i)^{{\bf U}_i}_{\bf s}\right)
                 \ar (\i{Body}_i)^{{\bf U}_i}_{\bf s}
$$
where {\bf s} is a tuple of precomputed terms of the same length as
${\bf U}_i$.  The infinitary rules with the consequent~$p({\bf r})$
contributed by this instance to $\tau_1\Gamma$ have the form
$$
\tau\left((\i{Body}_i)^{{\bf U}_i}_{\bf s}\right)\rar p({\bf r})
$$
where~{\bf s} satisfies the condition
${\bf r}\in[({\bf t}_i)^{{\bf U}_i}_{\bf s}]$.
If~$R_i$ is a choice rule
\beq
\{p({\bf t}_i)\} \ar \i{Body}_i
\eeq{ci}
then its instances have the form
$$
\{p\left(({\bf t}_i)^{{\bf U}_i}_{\bf s}\right)\}
                 \ar (\i{Body}_i)^{{\bf U}_i}_{\bf s}
$$
and the corresponding rules of~$\tau_1\Gamma$ with the consequent~$p({\bf r})$
have the form
$$
\tau\left((\i{Body}_i)^{{\bf U}_i}_{\bf s}\right)\land\neg\neg p({\bf r})
                                                           \rar p({\bf r}).
$$
Let~$G_i$ stand for
$\tau(\i{Body}_i)$
if~$R_i$ is a basic rule~(\ref{bi}), and for
$\tau(\i{Body}_i)\land\neg\neg p({\bf r})$
if~$R_i$ is a choice rule~(\ref{ci}). Using this notation, we can represent
formula~(\ref{dis1}) as
$$\bigvee_{i=1}^k\;
  \bigvee_{{\bf s}\,:\,{\bf r}\in[({\bf t}_i)^{{\bf U}_i}_{\bf s}]}
        (G_i)^{{\bf U}_i}_{\bf s}.$$
An interpretation~$\I$ satisfies this formula iff
\beq
\hbox{for some }i\in\{1,\dots,k\}\hbox{ and some {\bf s} such that }
{\bf r}\in[({\bf t}_i)^{{\bf U}_i}_{\bf s}],\ 
   \I\models (G_i)^{{\bf U}_i}_{\bf s}.
\eeq{dis1a}

On the other hand,~$F_i$ in disjunction~(\ref{dis2}) is
$${\bf V} \in {\bf t_i} \land \phi^{{\bf X}_i}(\i{Body}_i)$$
if~$R_i$ is a basic rule~(\ref{bi}), and
$${\bf V} \in {\bf t_i} \land \phi^{{\bf X}_i}(\i{Body}_i)\land p({\bf V})$$
if~$R_i$ is a choice rule~(\ref{ci}), where
${\bf X}_i$ is the list of local variables of rule~$R_i$.
Let~$H_i$ stand for
$\phi^{{\bf X}_i}(\i{Body}_i)$
if $R_i$ is~(\ref{bi}), and for
$\phi^{{\bf X}_i}(\i{Body}_i)\land p({\bf r})$
if $R_i$ is~(\ref{ci}).  Formula~(\ref{dis2}) can be written as
$$\bigvee_{i=1}^k\exists {\bf U}_i ({\bf r}\in {{\bf t}_i}\land H_i).$$
An intepretation~$\I$ satisfies this formula iff
\beq
\hbox{for some }i\in\{1,\dots,k\}\hbox{ and some {\bf s}},\quad
{\bf r}\in[({\bf t}_i)^{{\bf U}_i}_{\bf s}]\hbox{ and }
                \I\models (H_i)^{{\bf U}_i}_{\bf s}.
\eeq{dis2a}
Lemma~\ref{lem:phitau} shows that formulas
$(G_i)^{{\bf U}_i}_{\bf s}$ and $(H_i)^{{\bf U}_i}_{\bf s}$
are satisfied by the same interpretations.  Consequently
condition~(\ref{dis2a}) is equivalent to condition~(\ref{dis1a}).
\qed\medskip

\begin{lemma}
For any EG program~$\Gamma$, the infinitary formula~$\tau_2\Gamma$ is
satisfied by the same interpretations of the vocabulary of~$\Gamma$ as
the conjunction of the universal closures of the formula representations
of the constraints of~$\Gamma$.
\label{lem:cons}
\end{lemma}

\proof
We will show, for every constraint $\ar\i{Body}$ from~$\Gamma$, that the
universal closure of its formula representation $\phi(\ar{Body})$ is
satisfied by the same interpretations of the vocabulary of~$\Gamma$ as
the conjunction of the formulas
\beq
\neg\tau(\i{Body}^{\bf U}_{\bf r})
\eeq{consa}
for all tuples~{\bf r} of precomputed terms of the same length as the
tuple~{\bf U} of the global variables of $\ar\i{Body}$.  Recall that
$\phi(\ar{Body})$ is defined as $\neg\phi^{\bf X}(\i{Body})$, where~{\bf X} is
the list of local variables of $\ar\i{Body}$.  An interpretation~$\I$
satisfies the universal closure of this formula iff it satisfies the formulas
\beq
\neg\phi^{\bf X}(\i{Body}^{\bf U}_{\bf r})
\eeq{consb}
for all tuples~{\bf r} of precomputed terms of the same length as~{\bf U}.
By Lemma~\ref{lem:phitau}, formulas~(\ref{consa}) and~(\ref{consb}) are
satisfied by the same interpretations.
\qed

\subsection{Proof of Theorem~\ref{thm1}}

An interpretation~$\I$ is {\sl supported} by an infinitary program~$\Pi$ if
for each atom $A$ in~$\I$ there exists an infinitary formula~$F$ such that
$F\rar A$ is a rule of~$\Pi$ and $\I$ satisfies $F$. 
Every stable model of an infinitary program is supported by it
\cite[Lemma~B]{lif13a}.\footnote{See the long version of the paper,
{\tt http://www.cs.utexas.edu/users/ai-lab/?ltc}.}
It is easy to see that an interpretation~$\I$ satisfies the completion
of~$\Pi$ iff~$\I$ satisfies~$\Pi$ and is supported by~$\Pi$.  We conclude:
\begin{lemma}
Every stable model of an infinitary program satisfies its completion.
\label{forprop1}
\end{lemma}

To prove Theorem~\ref{thm1}, assume that~$\I$ is a stable model of an EG
program~$\Gamma$.  Then~$\I$ is a stable model of~$\tau_1\Gamma$, and~$\I$
satisfies~$\tau_2\Gamma$ (Lemma~\ref{lem:tau12}).  Consequently~$\I$
satisfies the completion of~$\tau_1\Gamma$ (Lemma~\ref{forprop1}).  It
follows that~$\I$ satisfies the completed definitions of all
predicate symbols occurring in~$\Gamma$ (Lemma~\ref{lem:comp}).  On the
other hand, since~$\I$ satisfies~$\tau_2\Gamma$, it satisfies also
the universal closures of the formula representations
of the constraints of~$\Gamma$ (Lemma~\ref{lem:cons}).
\qed

\subsection{Proof of Theorem~\ref{thm2}}

The proof of Theorem~\ref{thm2} below refers to the concept of a tight
infinitary program \cite{lif13a}.  We first define the set $\pnn(F)$ of
{\sl positive nonnegated atoms} of an infinitary formula~$F$ and
the set~$\nnn(F)$ of {\sl negative nonnegated atoms} of~$F$:
\begin{itemize}
\item $\pnn(\bot)=\emptyset$.
\item For any atom $A$, $\pnn(A)=\{A\}$.
\item $\pnn(\mathcal{H}^\land)=\pnn(\mathcal{H}^\lor)
       =\bigcup_{H\in\mathcal{H}}\pnn(H)$.
\item $\pnn(G\rar H)=\left\{\ba{ll}\emptyset & \mbox{if}\ H=\bot, \\
\nnn(G)\cup\pnn(H) & \hbox{otherwise.}\ea\right.$
\item $\nnn(\bot)=\emptyset$.
\item For any atom~$A$, $\nnn(A)=\emptyset$.
\item $\nnn(\mathcal{H}^\land)=\nnn(\mathcal{H}^\lor)
       =\bigcup_{H\in\mathcal{H}}\nnn(H)$.
\item $\nnn(G\rar H)=\left\{\ba{ll}\emptyset & \hbox{if}\ H=\bot, \\
\pnn(G)\cup\nnn(H) & \hbox{otherwise.}\ea\right.$
\end{itemize}

Let~$\Pi$ be an infinitary program, and~$\I$ an interpretation of its signature.
About atoms $A,B\in \I$ we say that $B$ is a {\sl parent of $A$
relative to $\Pi$ and $\I$} if there exists a formula~$F$ such that
$F\rar A$ is a rule of~$\Pi$, $\I$ satisfies $F$, and $B$ is
a positive nonnegated atom of~$F$.  We say that~$\Pi$
is {\sl tight on $\I$} if there is no infinite sequence $A_0, A_1, \ldots$ of
elements of~$\I$ such that for every~$i$, $A_{i+1}$ is a parent of $A_i$
relative to~$\Pi$ and~$\I$.

If an infinitary program~$\Pi$ is tight on an interpretation~$\I$ that
satisfies~$\Pi$ and is supported by~$\Pi$ then~$\I$ is a stable model
of~$\Pi$ \cite[Lemma~2]{lif13a}.  We conclude:

\begin{lemma} 
If an infinitary program~$\Pi$ is tight on an interpretation~$\I$ that
satisfies the completion of~$\Pi$ then~$\I$ is a stable model
of~$\Pi$.
\label{lem:ly}
\end{lemma}

\begin{lemma}
For any conjunction~{\bf C} of ground literals, ground comparisons, and
closed  aggregate expressions, if $p(t_1,\dots,t_n)$ is a positive
nonnegated atom of~$\tau{\bf C}$ then~$p/n$ occurs in a positive
literal or in an aggregate expression in~{\bf C}.
\label{lem:p2a}
\end{lemma}

\proof
Consider the conjunctive term~$C$ of {\bf C} such that $p(t_1,\dots,t_n)$
is a positive nonnegated atom of~$\tau C$.  It is clear from the
definition of~$\tau$ that~$p/n$ occurs in~$C$.  On the other hand,
the formulas obtained by applying~$\tau$ to negative literals and
comparisons have no positive nonnegated atoms.  Consequently~$C$ is either
a positive literal or an aggregate expression.
\qed

\begin{lemma}
If an EG program~$\Gamma$ is tight then $\tau_1\Gamma$ is tight on all
interpretations.
\label{lem:p2b}
\end{lemma}

\proof
Assume that $\tau_1\Gamma$ is not tight on an interpretation~$\I$, and
consider an infinite sequence $p_0({\bf t}_0),p_1({\bf t}_1),\dots$ of atoms
such that for every~$i$, $p_{i+1}({\bf t}_{i+1})$ is a parent of
$p_i({\bf t}_i)$ relative to~$\tau_1\Gamma$ and~$\I$.  We will show that
for every~$i$, the graph $G_{\tau_1\Gamma}$ has an edge from $p_i/n_i$ to
$p_{i+1}/n_{i+1}$, where~$n_i$ is the length of ${\bf t}_i$.  The the assertion
of the lemma will follow, because an infinite path  $p_0/n_0,p_1/n_1,\dots$
in the finite graph $G_{\tau_1\Gamma}$ is impossible if that graph is acyclic.

Consider a rule~$F_i\rar p_i({\bf t}_i)$ of $\tau_1\Gamma$ such that
$p_{i+1}({\bf t}_{i+1})$ is a positive nonnegated atom of~$F_i$.  This
rule has either the form~(\ref{basicinf}) or the form~(\ref{choiceinf}).
In both cases, $p_{i+1}({\bf t}_{i+1})$ is a positive nonnegated atom of
$\tau(\i{Body})$, and we can conclude, by Lemma~\ref{lem:p2a}, that
$p_{i+1}/n_{i+1}$ occurs in a positive literal or in an aggregate
expression in \i{Body}.  It remains to observe that \i{Body} is the body of
an instance of a rule of $\tau_1\Gamma$ that contains $t_i/n_i$ in the head.
\qed

\medskip\noindent{\bf Proof of Theorem~\ref{thm2}$\ $}
Let $\Gamma$ be a finite tight EG program.  Given Theorem~1, we only
need to establish the ``if'' direction of Theorem~\ref{thm2}: if an
interpretation of the vocabulary of~$\Gamma$ satisifies the completion
of~$\Gamma$ then it is a stable model of~$\Gamma$.

Let~$\I$ be an interpretation of the vocabulary of~$\Gamma$ that satisfies
the completion of~$\Gamma$.  Then~$\I$ satisfies the completion
of~$\tau_1\Gamma$ (Lemma~\ref{lem:comp}).  But~$\tau_1\Gamma$ is tight
on~$\I$ (Lemma~\ref{lem:p2b}); consequently~$\I$ is a stable model
of~$\tau_1\Gamma$ (Lemma~\ref{lem:ly}).  On the other hand,
$\I$ satisfies~$\tau_2\Gamma$ (Lemma~\ref{lem:cons}).  It follows
that~$\I$ is a stable model of~$\Gamma$ (Lemma~\ref{lem:tau12}).
\qed


\begin{thebibliography}{}

\bibitem[\protect\citeauthoryear{Clark}{Clark}{1978}]{cla78}
{\sc Clark, K.} 1978.
\newblock Negation as failure.
\newblock In {\em Logic and Data Bases}, {H.~Gallaire} {and} {J.~Minker}, Eds.
  Plenum Press, New York, 293--322.

\bibitem[\protect\citeauthoryear{Erdem and Lifschitz}{Erdem and
  Lifschitz}{2003}]{erd03}
{\sc Erdem, E.} {\sc and} {\sc Lifschitz, V.} 2003.
\newblock Tight logic programs.
\newblock {\em Theory and Practice of Logic Programming\/}~{\em 3}, 499--518.

\bibitem[\protect\citeauthoryear{Fages}{Fages}{1994}]{fag94}
{\sc Fages, F.} 1994.
\newblock Consistency of {C}lark's completion and existence of stable models.
\newblock {\em Journal of Methods of Logic in Computer Science\/}~{\em 1},
  51--60.

\bibitem[\protect\citeauthoryear{Ferraris}{Ferraris}{2005}]{fer05}
{\sc Ferraris, P.} 2005.
\newblock Answer sets for propositional theories.
\newblock In {\em Proceedings of International Conference on Logic Programming
  and Nonmonotonic Reasoning ({LPNMR})}. 119--131.

\bibitem[\protect\citeauthoryear{Ferraris, Lee, and Lifschitz}{Ferraris
  et~al\mbox{.}}{2011}]{fer09}
{\sc Ferraris, P.}, {\sc Lee, J.}, {\sc and} {\sc Lifschitz, V.} 2011.
\newblock Stable models and circumscription.
\newblock {\em Artificial Intelligence\/}~{\em 175}, 236--263.

\bibitem[\protect\citeauthoryear{Ferraris and Lifschitz}{Ferraris and
  Lifschitz}{2005}]{fer05e}
{\sc Ferraris, P.} {\sc and} {\sc Lifschitz, V.} 2005.
\newblock Mathematical foundations of answer set programming.
\newblock In {\em We Will Show Them! Essays in Honour of Dov Gabbay}. King's
  College Publications, 615--664.

\bibitem[\protect\citeauthoryear{Ferraris and Lifschitz}{Ferraris and
  Lifschitz}{2010}]{fer10a}
{\sc Ferraris, P.} {\sc and} {\sc Lifschitz, V.} 2010.
\newblock The stable model semantics for first-order formulas with aggregates.
\newblock In {\em Proceedings of International Workshop on Nonmonotonic
  Reasoning (NMR)}.

\bibitem[\protect\citeauthoryear{Gebser, Harrison, Kaminski, Lifschitz, and
  Schaub}{Gebser et~al\mbox{.}}{2015}]{geb15}
{\sc Gebser, M.}, {\sc Harrison, A.}, {\sc Kaminski, R.}, {\sc Lifschitz, V.},
  {\sc and} {\sc Schaub, T.} 2015.
\newblock Abstract {G}ringo.
\newblock {\em Theory and Practice of Logic Programming\/}~{\em 15}, 449--463.

\bibitem[\protect\citeauthoryear{Gurevich and Shelah}{Gurevich and
  Shelah}{1986}]{gur86}
{\sc Gurevich, Y.} {\sc and} {\sc Shelah, S.} 1986.
\newblock Fixed-point extensions of first-order logic.
\newblock {\em Annals of Pure and Applied Logic\/}~{\em 32}, 265--280.

\bibitem[\protect\citeauthoryear{Harrison, Lifschitz, Pearce, and
  Valverde}{Harrison et~al\mbox{.}}{2015}]{har15a}
{\sc Harrison, A.}, {\sc Lifschitz, V.}, {\sc Pearce, D.}, {\sc and} {\sc
  Valverde, A.} 2015.
\newblock Infinitary equilibrium logic and strong equivalence.
\newblock In {\em Proceedings of International Conference on Logic Programming
  and Nonmonotonic Reasoning ({LPNMR})}. 398--410.

\bibitem[\protect\citeauthoryear{Harrison, Lifschitz, and Yang}{Harrison
  et~al\mbox{.}}{2014}]{har14a}
{\sc Harrison, A.}, {\sc Lifschitz, V.}, {\sc and} {\sc Yang, F.} 2014.
\newblock The semantics of {G}ringo and infinitary propositional formulas.
\newblock In {\em Proceedings of International Conference on Principles of
  Knowledge Representation and Reasoning (KR)}.

\bibitem[\protect\citeauthoryear{Lee and Meng}{Lee and Meng}{2009}]{lee09}
{\sc Lee, J.} {\sc and} {\sc Meng, Y.} 2009.
\newblock On reductive semantics of aggregates in answer set programming.
\newblock In {\em Procedings of International Conference on Logic Programming
  and Nonmonotonic Reasoning ({LPNMR})}. 182--195.

\bibitem[\protect\citeauthoryear{Lee and Meng}{Lee and Meng}{2012}]{lee12a}
{\sc Lee, J.} {\sc and} {\sc Meng, Y.} 2012.
\newblock Stable models of formulas with generalized quantifiers.
\newblock In {\em Working Notes of the 14th International Workshop on
  Non-Monotonic Reasoning (NMR)}.

\bibitem[\protect\citeauthoryear{Lifschitz and Yang}{Lifschitz and
  Yang}{2013}]{lif13a}
{\sc Lifschitz, V.} {\sc and} {\sc Yang, F.} 2013.
\newblock Lloyd-{T}opor completion and general stable models.
\newblock {\em Theory and Practice of Logic Programming\/}~{\em 13,\/}~4--5.

\bibitem[\protect\citeauthoryear{Truszczynski}{Truszczynski}{2012}]{tru12}
{\sc Truszczynski, M.} 2012.
\newblock Connecting first-order {ASP} and the logic {FO(ID)} through reducts.
\newblock In {\em Correct Reasoning: Essays on Logic-Based AI in Honor of
  Vladimir Lifschitz}, {E.~Erdem}, {J.~Lee}, {Y.~Lierler}, {and} {D.~Pearce},
  Eds. Springer, 543--559.

\end{thebibliography}
\end{document}